\documentclass{PoS}

\newcommand{\G}{\Gamma}

\newcommand{\sT}{\sigma_{\rm T}}
\newcommand{\p}{^\prime}
\newcommand{\e}{\epsilon}
\newcommand{\g}{\gamma}
\newcommand{\gp}{\gamma^{\prime}}

\newcommand{\ep}{\epsilon^\prime}

\newcommand{\up}{u^\prime}

\newcommand{\psim}{\lower.5ex\hbox{$\; \buildrel \propto \over\sim \;$}}
\newcommand{\lbar}{\lower.0ex\hbox{$\; \buildrel
{\lower0.0ex \hbox{-}} \over\lambda  \;$}}

\newcommand{\ph}{\mathrm{ph}}
\newcommand{\cm}{\mathrm{cm}}

\newcommand{\erg}{\mathrm{erg}}
\newcommand{\eV}{\mathrm{eV}}

\newcommand{\MeV}{\mathrm{MeV}}
\newcommand{\GeV}{\mathrm{GeV}}

\newcommand{\s}{\mathrm{s}}
\newcommand{\dday}{\mathrm{day}}
\newcommand{\Hz}{\mathrm{Hz}}
\newcommand{\pc}{\mathrm{pc}}

\newcommand{\Gauss}{\mathrm{G}}

\newcommand{\apj}{ApJ}
\newcommand{\aj}{AJ}
\newcommand{\apjl}{ApJL}
\newcommand{\apjs}{ApJS}
\newcommand{\aap}{A\&A}
\newcommand{\mnras}{MNRAS}
\newcommand{\pasp}{PASP}
\newcommand{\aaps}{A\&AS}

\newcommand{\nat}{Nature}
\newcommand{\prd}{Phys.~Rev.~D}

\newcommand{\jcap}{JCAP}

\title{Modeling {\em Fermi} Large Area Telescope and Multiwavelength
Data from Blazars}

\ShortTitle{Modeling Multiwavelength Data from Blazars}

\author{\speaker{Justin D.\ Finke}\\
        U.S. Naval Research Laboratory, Code 7653, 4555 Overlook Ave.\ SW, Washington, DC 20375 USA\\
        E-mail: \email{justin.finke@nrl.navy.mil}}

\abstract{

Blazars are active galactic nuclei with relativistic jets pointed at
the Earth, making them extremely bright at essentially all
wavelengths, from radio to gamma rays.  I will review the modeling of
this broadband spectral energy distributions of these objects, and
what we have learned, with a focus on gamma rays.}

\FullConference{3rd Annual Conference on High Energy Astrophysics in Southern Africa  -HEASA2015,\\
		18-20 June 2015\\
		University of Johannesburg, Auckland Park, South Africa}

\begin{document}

\section{Introduction}

In this proceeding, I review results of multi-wavelength modeling of
{\em Fermi} LAT-detected blazars.  It is an update of my previous
short review from the {\em Fourth Fermi Symposium}
\cite{finke13_review}.  The remainder of this section is devoted to a
basic introduction to blazars.  In Section \ref{population} I discuss
some results of population studies of blazars, with an emphasis on the
blazar sequence.  Section \ref{curvature} is devoted to a discussion
of the origin of the curvature in the $\g$-ray spectra of some
blazars, one of the major discoveries of the {\em Fermi}-LAT that does
not currently have a good explanation.  Another major open problem in
blazar physics is the question of where in the jet the $\g$-ray flares
are produced, and this is the topic of Section \ref{location}.  One of
the defining characteristics of blazars is that they are highly
variable, and so this is discussed in Section \ref{variability}.
Finally, in Section \ref{hadronic}, a brief description of hadronic
models for $\g$-ray emission in blazars is given.

\subsection{The {\em Fermi Gamma Ray Space Telescope}}

The {\em Fermi Gamma Ray Space Telescope}, launched from Cape
Canaveral, Florida, USA, on 2008 June 11 has two instruments, the
Large Area Telescope (LAT), and the Gamma-Ray Burst Monitor.  Only the
former will be discussed here.  The LAT is a pair conversion telescope
sensitive to photons in the 20 MeV to 300 GeV range with about
0.8$^\circ$ resolution for a single photon \cite{atwood09}.  It sees
about 20\% of the sky at any one time, and surveys the entire sky
every 3 hours.  Eighty-five percent of the 2023 associated or
identified sources in the most recent {\em Fermi} LAT Catalog, the
3FGL, are {\em blazars} \cite{acero15_3fgl}.

% 1717 blazars in 3FGL

\subsection{Blazars and Their Basic Properties}

It is thought that almost all (if not all) galaxies have supermassive
black holes ($M_{BH} \sim 10^6-10^{10}\ M_{\odot}$) at their centers.
About 20\% of these have bright nuclei powered by accretion onto the
supermassive black hole, and are known as active galactic nuclei (AGN;
\cite{kauffman03}). About 15-20\% of these are radio loud
\cite{kellermann89}, with their radio emission being due to a
relativistic jet.  Quasars with flat radio spectra are thought to have
their jet pointed towards the Earth.  The flat spectrum radio quasars
and BL Lacertae objects together are known as blazars.  They are
characterized by multi-wavelength highly variable (timescales as short
as minutes in some cases) nonthermal emission at essentially all
frequencies, from radio to $\g$-rays.  They are thought to be
misaligned radio galaxies, which have extended jets not pointed
towards our line of sight, terminating in radio lobes at $\sim0.1$--1
Mpc distances from the black hole.

Individual components in blazar jets have been resolved with radio
very long baseline interferometry (VLBI).  There is considerable
evidence that these individual components are traveling at a large
fraction $\beta$ of the speed of light $c$ (giving them a bulk Lorentz
factors $\G = (1-\beta)^{-1/2}$) at small angles ($\theta$) to our
line of sight.  This evidence includes:
\begin{itemize}
\item Extremely high radio surface brightnesses, which
would require extreme energy densities if produced by synchrotron
from stationary sources (e.g.,\cite{jones74a,jones74b}).
\item On the milliarcsecond scale, jet components can be resolved with
radio VLBI (e.g.\ \cite{lister09}) and they often appear to have
apparent speeds $v_{app}=\beta_{app}c>c$, which can be explained by
blobs moving at near the speed of light ($\beta\sim1$, $\G \gg 1$)
with angles close to the line of sight ($\theta\ll 1$).  In this case
their apparent speed is\footnote{I will use the notation $A_x=10^x
A$ with Gaussian/cgs units unless otherwise stated.}
\begin{equation}
\beta_{app} = \frac{\beta \sin\theta}{1-\beta\cos\theta} =
\delta\G\beta\sin\theta \approx 10\ \delta^2_1(\G/\delta)\theta_{-1}
%\approx \frac{2\G^2\theta}{1+\G^2\theta^2}
\end{equation}
where $\delta=[\G(1-\beta\cos\theta)]^{-1}$.
\item The detection of rapid $\gamma$-ray flares from blazars implies
the blob must be moving with $\G\gg1$ in order for the $\g$ rays to
escape and avoid $\g\g$ attenuation (e.g., \cite{dondi95}).
\end{itemize}

This geometry leads to a number of observational consequences.
Events in the blob will appear in ``fast forward'', so that the
observer sees a time interval $t=(1+z)t\p/\delta$ where $t\p$ is some
time interval in the frame comoving with the blob\footnote{In general,
primed quantities will refer to this comoving frame and unprimed
quantities will refer to the observer frame.} and $z$ is the
cosmological redshift.  Light-travel time effects imply that a
spherical blob must have a radius
\begin{equation}
\label{Rblob}
R\p_b \le \frac{\delta c t_v}{1+z} = 3\times10^{15}\ \delta_1
t_{v,4}(1+z)^{-1}\ \cm
\end{equation}
where $t_v$ is the variability time, approximately the time it takes
the flux to change by a factor of 2.  The observed photon energy is
``blueshifted'' $\e=\delta\ep/(1+z)$, and the flux is
``Doppler-boosting'' so that if the comoving frame $\nu F_\nu$
synchrotron flux is $f\p_{sy}$, the observer sees
$f_{sy}=\delta^4f\p_{sy}$.  A blob with $\delta=10$ appears
$\sim10^4$ times brighter than it would if it was stationary.

\subsection{Blazar Spectral Energy Distributions (SEDs)}

\begin{figure}
%\vspace{4.mm}
\vspace{8.mm}
\includegraphics[width=0.6\textwidth]{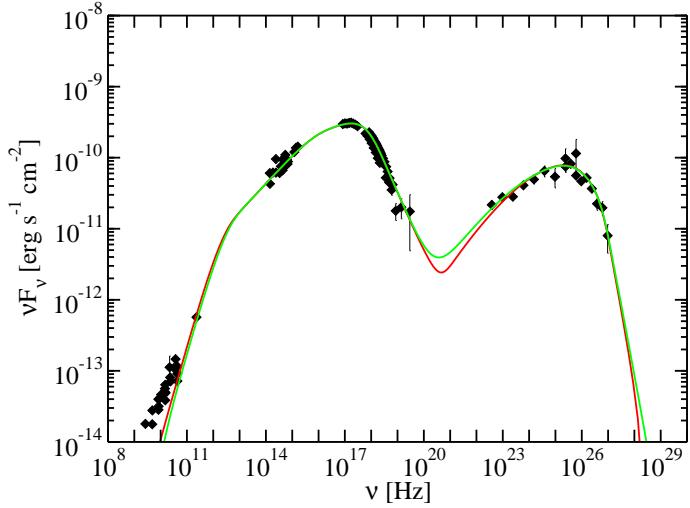}
\caption{The SED of the HSP BL Lac object Mrk 421, with
$\nu_{pk}^{sy}\approx 10^{17.3}$\ Hz.  Figure taken from
\cite{abdo11_mrk421}.}
\label{mrk421sed}
\end{figure}

Blazars' spectral energy distributions (SEDs) have two components
(Figure \ref{mrk421sed}): a low-frequency component, peaking in the
infrared to X-rays, almost certainly from synchrotron emission; and a
high-frequency component, peaking at $\g$-ray energies.  The origin of
the $\g$-ray component is not completely certain, but it is probably
due to the Compton-scattering of a seed photon source which could be
either the synchrotron photons themselves, in which case it is known
as synchrotron self-Compton (SSC); or they could be external to the
jet, often known as external Compton (EC).  The source of seed photons
for EC could be the accretion disk \cite{dermer93,dermer02},
broad-line region (BLR; \cite{sikora94}), dust torus \cite{kataoka99},
or even other components of the jet \cite{ghisellini05,macdonald15}.
The broadband emission is usually modeled as being from a single
component (a ``one zone'' model'') where all the emission is generated
by a single population of electrons, except for the radio which is
thought to be from the superposition of several self-absorbed
components \cite{konigl81}.  This can be justified to some extent by
correlated variability between several wavebands (e.g.,
\cite{fossati08,aharonian09_2155,bonning09,chatterjee12,rani13,cohen14}).

Aside from Compton scattering, $\g$ rays could also be produced by
protons co-accelerated in the jet with the electrons.  This is
motivated by the fact that blazars are a leading contender to be the
origin of ultra-high energy cosmic rays (UHECRs; e.g.,
\cite{dermer10}).  Protons could produce $\g$ rays directly through
synchrotron emission or through the synchrotron emission of child
particles produced through p$\g$ interactions
\cite{mannheim92,mannheim93,muecke01,muecke03}.

\vspace{1.2mm}
%\clearpage

\subsection{Classification of Blazars}

Blazars are classified in two basic ways: based on their emission line
properties, and based on the peak frequency of their synchrotron
component.

The classification of blazars based on their emission lines has
changed several times, but basically, blazars are classified as {\em
BL Lac objects} if the equivalent width of their broad lines are less
than a certain value, and as {\em Flat Spectrum Radio Quasars (FSRQs)}
otherwise, taking into account the strength of the Ca H \& K break
\cite{stocke91,marcha96} and possibly narrow emission lines
\cite{landt04}.  A different, more physical transition based on
luminosity was suggested by Ghisellini et al.\
\cite{ghisellini11_transition}.  They proposed that those with broad
line luminosity as a fraction of Eddington
$L_{BLR}/L_{Edd}<5\times10^{-4}$ be considered BL Lac objects, while
those with $L_{BLR}/L_{Edd}>5\times10^{-4}$ be considered FSRQs.  BL
Lacs are thought to be associated with Fanaroff-Riley (FR;
\cite{fanaroff74}) type I radio galaxies, and FSRQs with FR type II
radio galaxies \cite{urry95}, although exceptions are known to exist
(e.g., \cite{landt08}).

Blazars are also classified based on the frequency of their
synchrotron peak, $\nu_{pk}^{sy}$.  In the latest incarnation of this,
they are considered low-synchrotron peaked (LSP) if
$\nu_{pk}^{sy}<10^{14}$\ Hz; intermediate-synchrotron peaked (ISP) if
$10^{14}\ \Hz\ < \nu_{pk}^{sy} < 10^{15}\ \Hz$; and high-synchrotron
peaked if $10^{15}< \nu_{pk}^{sy}\ \Hz$ \cite{abdo10_sed}.  Almost all
FSRQs are LSPs \cite{ackermann11_2lac,ackermann15_3lac}.  Blazars are
highly variable, and their peak frequency can change over time (e.g.,
\cite{ghisellini13}).  This means that the classification of a blazar
can depend on the epoch it was observed.

\section{Population Studies}
\label{population}

\subsection{Blazar Sequence}

A useful tool in stellar astrophysics is the Hertzsprung-Russell
diagram, which describes the luminosity and the optical spectral type
(related to temperature and color) of stars and includes the very
prominent main sequence, on which stars spend a large fraction of
their lifetimes.  Due to its success in understanding stars, it is
tempting to look for a similar diagram for blazars.  A ``blazar main
sequence'' or ``blazar sequence'' was suggested by Fossati et al.\
\cite{fossati98}, combining three samples of blazars
(\cite{kuhr81,wall85,elvis92}).  They found three parameters that
appeared to be well-correlated with $\nu_{pk}^{sy}$: the 5 GHz radio
luminosity, the luminosity at the peak of the synchrotron component
($L_{pk}^{sy}$), and the ``$\gamma$-ray dominance'', i.e., the ratio
of of the $\g$-ray luminosity ($L_\g$; as measured by EGRET) and the
peak luminosity of the synchrotron component ($L_{pk}^{sy}$).  The
existence of these correlations in subsequent studies is a matter of
some debate \cite{padovani03,nieppola06,chen11_sequence,finke13}.

Nevertheless, Ghisellini et al.\ \cite{ghisellini98} suggested a
physical explanation for these correlations.  For nonthermal electrons
accelerated as power-laws and allowed to escape a region of size
$R^\prime$ and cool through synchrotron and Compton losses, a
``cooling break'' is found in the electron distribution at electron
Lorentz factor given by
\begin{eqnarray}
\label{gc}
\g_c^\prime = \frac{ 3 m_e c^2 }{ 4 c \sT u_{tot}^{\prime} t^\prime_{esc} }\ ,
\end{eqnarray}
where $m_e=9.1\times10^{28}$\ g is the electron mass,
$\sT=6.65\times10^{-25}\ \cm$ is the Thomson cross section,
$t^\prime_{esc}\cong R^\prime/c$ is the escape timescale, and
$u_{tot}^\prime$ is the total energy density in the frame of the
relativistic blob, given by the sum of the Poynting flux
($u_B^\prime$), synchrotron ($u_{sy}^\prime$), and external radiation
field ($u^\prime_{ext} \cong \G^2 u_{ext}$) energy densities.  
The cooling Lorentz factor $\gp_c$ is associated with a peak in
the synchrotron spectrum of the source in a $\nu F_\nu$ representation
observed at frequency
\begin{eqnarray}
\label{nupk}
\nu_{pk}^{sy} = 3.7\times10^6\ \g_c^{\prime 2}\ \left(\frac{B}{\Gauss}\right)\ 
\frac{\delta}{1+z}\ \Hz
%\\ \nonumber
%= 3.6\times10^{14}\ \Hz\ B_0\ \delta_{1}\ (1+z)^{-1}
%\ R_{15.5}^{\prime -2}\ u_{tot,-1}^{\prime -2}\ .
\end{eqnarray}
(e.g., \cite{tavecchio98}) where $B$ is the magnetic field in the
blob.  For objects that have weak external radiation fields so that
$\up_B \gg \up_{ext}$, and neglecting $\up_{sy}$, Equations (\ref{gc})
and (\ref{nupk}) give
\begin{eqnarray}
\nu_{pk}^{sy} \cong 2.2\times10^{15}\ B_0^{-3}\ \delta_1\ (1+z)^{-1}\ 
\ R_{15.5}^{\prime -2}\ \Hz\ .
\end{eqnarray}
These objects are HSPs.  Objects with a strong external radiation
fields from the broad line region (BLR) which dominate over $\up_B$
and $\up_{sy}$, have peak synchrotron frequencies given by
\begin{eqnarray}
\nu_{pk}^{sy} \cong 3.2\times10^{12}\ \ B_0\ \delta_1^{-3}\ (\delta/\G)^4\ (1+z)^{-1}\ 
\ R_{15.5}^{\prime -2}\ u_{ext,-2}^{-2}\ \Hz\ .
\end{eqnarray}
These objects are LSPs.  Thus far, all HSP blazars
($\nu_{pk}^{sy}>10^{15}$\ Hz) discovered have been BL Lac objects with
weak if any broad emission lines, while almost all FSRQs, with strong
emission lines, have been LSPs ($\nu_{pk}^{sy}<10^{14}$\ Hz).
However, there are a significant number of BL Lacs which are LSPs.
Objects with stronger line emission would also be expected to have
greater $\g$-ray dominances, due to scattering of the external
radiation field.  Ghisellini et al.\ \cite{ghisellini98} thus
predicted a sequence of blazars, from low power, high peaked, low
$\g$-ray dominance, lineless objects, and as the external radiation
field increases, to low peaked, high $\g$-ray dominance objects with
strong broad emission lines.

In all of the {\em Fermi}-LAT AGN catalogs, a correlation between
$\gamma$-ray spectral index ($\G_\g$) and $\nu_{pk}^{sy}$ has been
found
\cite{abdo09_lbas,abdo10_sed,abdo10_1lac,ackermann11_2lac,ackermann15_3lac}.
This correlation was explained by Dermer et al.\ \cite{dermer15}.  As
the synchrotron peak moves to higher frequencies, Compton peak will
also move to higher frequencies, resulting in harder $\g$-ray spectra,
for a jet blob with a log-parabola electron distribution that is in
near-equipartition with the Poynting flux.  This could also represent
a diagram around which one could build a ``blazar main sequence''.

\subsection{Alternative Explanations for the Blazar Sequence}

\begin{figure}
%\vspace{4.mm}
\vspace{8.mm}
\includegraphics[width=0.6\textwidth]{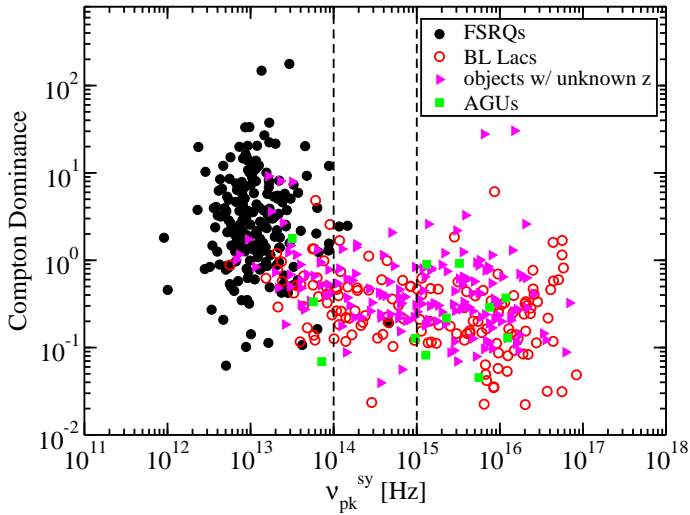}
\caption{Compton dominance (i.e., $L_{pk}^C/L_{pk}^{sy}$) versus peak
synchrotron frequency.  Filled circles represent FSRQs, empty circles
represent BL Lacs, and filled squares represent objects which do not
have an unambiguous classification.  Rightward-pointing triangles
represent BL Lacs with unknown redshifts, for which $\nu_{pk}^{sy}$ is
a lower limit.  Figure taken from \cite{finke13}.}
\label{CD}
\end{figure}

An alternative explanation for the correlations found by Fossati et
al.\ was given by Giommi and collaborators
\cite{giommi02,giommi05,giommi12_selection}. They propose that the
sequence is a result of a selection effect: luminous blazars with high
synchrotron peaks have their spectral lines totally swamped by the
nonthermal continuum, making a redshift measurement impossible.
Without a redshift, it is not possible to determine their
luminosities, and so they are not included in statistical tests that
measure a correlation between luminosity and $\nu_{pk}^{sy}$.  There
is some evidence for this \cite{rau12,padovani12}, although see Meyer
et al.\ \cite{meyer12} and Ghisellini et al.\ \cite{ghisellini12}.
However, an intrinsic quantity that can be determined without redshift
could be used in place of luminosity.  The blazar sequence as
originally found by Fossati et al.\ \cite{fossati98} included the
$\g$-ray dominance.  Note also that $\nu_{pk}^{sy}$ is only weakly
dependent on redshift, by a factor $(1+z)$, i.e., a factor of a few.
The {\em Fermi}-LAT allows the $\g$-ray dominance and the Compton
dominance ($A_C\equiv L_{pk}^C / L_{pk}^{sy}$ where $L_{pk}^C$ is the
luminosity at the Compton peak) to be found for more blazars than in
the EGRET era.  A plot of $A_C$ versus $\nu_{pk}^{sy}$ is shown in
Fig.\ \ref{CD}, from a subset of sources in the second LAT AGN catalog
\cite{ackermann11_2lac}, including sources that do not have known
redshifts.  A correlation clearly exists, as shown by the Spearman and
Kendall tests \cite{finke13}, so that this aspect of the blazar
sequence does seem to have a physical origin.

\vspace{1.2mm}
%\clearpage

Meyer et al.\ \cite{meyer11} proposed another physical scenario, based
on updated data from a number of sources.  In their scenario, the
difference between BL Lacs and FSRQs is the former have jet structure
with velocity (or Lorentz factor) gradients, either perpendicular or
parallel to the direction of motion, while FSRQs do not have these
gradients. They have a single Lorentz factor for the entire jet, or at
least the radiatively important parts.  There is indeed evidence for
different Lorentz factors in BL Lacs and FRIs (e.g.,
\cite{chiaberge00,chiaberge01,abdo10_cena}).  The lack of $\g$-ray
detected FRIIs hints that FRIIs/FSRQs do not share this jet structure
\cite{grandi12}, although, see \cite{boett09_decel} for evidence of jet
deceleration in an FSRQ.  The scenario of \cite{meyer11} predicts that
there should be no radio galaxies with $\nu_{pk}^{sy}\gtrsim10^{15}$\
Hz, since these will be only the most aligned BL Lac sources.
However, radio galaxies have recently been found with peaks this high
\cite{fukazawa15}, so that the scenario of Meyer et al.\ may need
revision.

\subsection{Blazar Evolution}

The LAT $\g$-ray luminosity functions, luminosity densities, and
number densities, and their evolution with redshift have been explored
by M.\ Ajello and collaborators \cite{ajello12,ajello14}.  As one goes
to higher redshift, the number density of FSRQs increases up to
$z\approx 0.5$, where it begins to decrease.  By contrast, the number
density of HSP BL Lacs always decreases as one goes to higher $z$,
with significantly steeper decrease between $z\approx 0.0-0.5$.
Ajello et al.\ suggested that this may be due to FSRQs turning into BL
Lac objects at redshifts $z\lesssim0.5$.  This provides some evidence
for an evolutionary scenario for the blazar sequence, as outlined by
B\"ottcher \& Dermer \cite{boett02_seq} and Cavaliere \& D'Elia
\cite{caval02}. In this scenario, FSRQs are younger objects and have
substantial circum-nuclear material making up the accretion disk and
broad-line region (BLR), which is slowly accreted onto the black hole.
As the circum-nuclear material accretes, the broad emission lines
decrease, and the accretion rate decreases, and the sources become
older BL Lac objects.

\section{Gamma-Ray Spectral Curvature}
\label{curvature}

After the launch of {\em Fermi}, while the spacecraft was still in its
post-launch commissioning and checkout phase, the FSRQ 3C~454.3 was
detected by the LAT in an extreme bright state \cite{tosti08_atel}.
The source reached a flux of $F(>100\ \MeV) > 10^{-5}\ \ph\ \cm^{-2}\
\s^{-1}$ and its spectrum showed an obvious curvature (i.e., a
deviation from a single power-law), which was best-fit by a broken
power-law \cite{abdo09_3c454.3} with break energy $\sim 2$\ GeV.  This
source flared on several more occasions
\cite{ackermann10_3c454.3,abdo11_3c454.3}, exhibiting a spectral break
(or spectral curvature) during bright states.  The energy of the break
varied by no more than a factor of $\sim 3$, while the flux varied by
as much as a factor of 10 \cite{abdo11_3c454.3}.  This spectral
curvature has been found in other blazars as well, although a broken
power-law is not always preferred over a log-parabola fit, which has
one less free parameter \cite{abdo10_latsed}.  The spectral break
(curvature) is not always present in flaring states of 3C 454.3 and
other blazars \cite{pacciani14}.  The cause of the curvature is not
clear but there are several possible explanations, which are briefly
reviewed below.

\subsection{Combination of Several Scattering Components}

It was noted that, based on the shape of the optical and $\g$-ray
spectra, the Compton scattering of more than one seed photon source
was needed to explain the overall spectral energy distribution (SED)
of 3C~454.3 \cite{finke10_3c454}.  The particularly soft spectra above
the break requires that this scattering be done in the Klein-Nishina
(KN) regime.  The FSRQ PKS 1424$-$418 had an unusual (although not
statistically significant) ``upturn'' in its LAT spectrum that was
modeled as the combination of several scattering components by Buson
et al.\ \cite{buson14}.  In order for this scenario to be viable, the
$\g$-ray emitting region must be within the BLR.

\subsection{Compton Scattering of BLR Ly$\alpha$ Photons} 

For the scattering of Ly$\alpha$ photons ($E_*=10.2$\ eV), the KN
regime will emerge at energies above
\begin{eqnarray}
E_{KN} \approx 1.2\ (E_*/10.2\ \eV)^{-1}\ \GeV\ ,
\end{eqnarray}
approximately in agreement with the observed break energy
\cite{ackermann10_3c454.3}.  Fits with this model using power-law
electron distributions failed to reproduce the observed LAT spectra
\cite{ackermann10_3c454.3}; however, fits using a log-parabola
electron distribution were able to reproduce the $\g$-rays
\cite{cerruti13}.  This model would also require the $\g$-ray emitting
region to be within the BLR.

\subsection{Curvature in the electron distribution}
\label{electron_curvature}

\begin{figure}
\vspace{8.mm}
\includegraphics[width=0.6\textwidth]{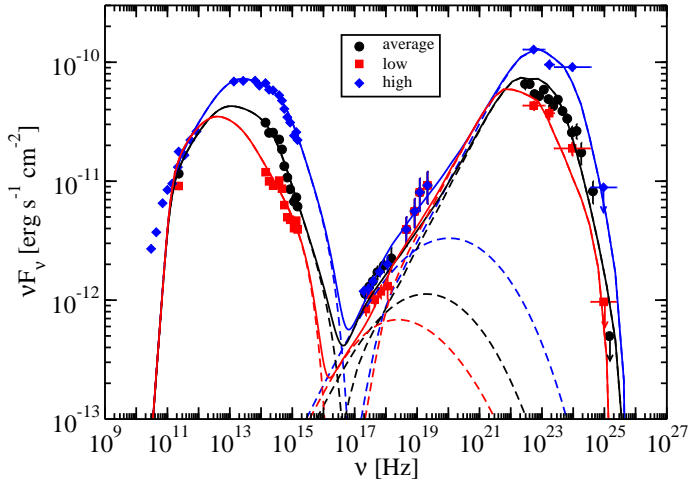}
\caption{SED of the FSRQ PKS~0537$-$441 \cite{dammando13_0537}.  The
spectral curvature in the IR/optical in the high state could indicate
the cause of the $\g$-ray curvature is the result of curvature in the
electron distribution.  }
\label{0537sed}
\end{figure}

Abdo et al.\ \cite{abdo09_3c454.3} suggested that if there is
curvature in the electron distribution that produces the $\g$-rays,
presumably from Compton scattering, this would naturally be reflected
in the LAT spectrum as well.  In this scenario, one would expect the
curvature in the electron distribution to cause a curvature in the
synchrotron emission from the same electrons, which would appear in
the IR/optical.  Indeed, observations of PKS 0537$-$441 do show this
curvature (see Fig.\ \ref{0537sed}; \cite{dammando13_0537}).  This
explanation would not require scattering to take place in the BLR, as
dust torus photons could be the seed photon source for scattering.

\subsection{Photoabsorption in the Broad Line Region}
\label{BLRabsorption}

Poutanen \& Stern \cite{poutanen10} modeled the breaks at a few GeV
with $\g\g$ absorption features from interactions of the $\g$ rays
with He II Ly$\alpha$ and recombination photons (54.4 eV and 40.8 eV,
respectively) from the BLR.  The existence of these breaks was
disputed by Harris et al.\ \cite{harris12}.  A few years later, with
an updated analysis including more LAT data (almost 5 years) and
updated instrument response functions, Stern \& Poutanen
\cite{stern14} found instead that the absorption features were at
$\sim10$\ GeV, with they interpreted as being from $\g\g$ absorption
by H Lyman continuum photons (13.6 eV).  An absorption feature at this
energy was actually predicted before the launch of {\em Fermi} by
Reimer \cite{reimer07}.  The $\g\g$ absorption optical depths implied
by the fits to the LAT data by Stern \& Poutanen \cite{stern14} were
quite a bit lower than expected, $\tau_{\g\g}\sim 2-5$, compared to
$\tau_{\g\g}\approx 100$ that one would expect if the $\g$ rays were
produced deep inside the BLR.  They proposed several possible
explanations for this: the size of the Lyman continuum BLR is larger
than expected; a flattened geometry for the BLR, leading to much of
the BLR photons being at an unfavorable geometry for $\g\g$
absorption; and finally, there may be several places along the jet
where $\g$ rays are emitted, some inside and some outside the BLR, and
all of these flares contribute to the $\g$-ray spectrum integrated
over several years.  The $\g$ rays from inside could be ``diluted'' by
photons from outside, creating the illusion of less absorption.

\subsection{Conversion to Axion-Like Particles}

Axion-like particles (ALPs) have been proposed in several predictions
of standard model extensions (e.g., \cite{ringwald12}).  It may be
possible for photons to transform into ALPs and back again in the
presence of a magnetic field.  It has been shown that this could fit
the spectral curvature seen in the $\g$-ray spectrum of 3C 454.3
\cite{mena13}, as well as allow photons to escape the BLR that
otherwise might not be able to do so \cite{tavecchio12,mena13}.

\section{Location of Gamma-Ray Emitting Region}
\label{location}

In recent years, there has been an intense debate as to the location
of the $\g$-ray emitting region in FSRQs, with no consensus yet
reached.  There are two main options: inside the BLR, within $\sim
0.1$\ pc from the black hole, where BLR photons are the likely seed
photon source for Compton scattering, and $\gtrsim 1$\ pc, where the
dust torus is the likely source of seed photons.  These possibilities
are explored below, with some ways to distinguish them based on
$\g$-ray light curves and gravitational lensing.

\subsection{Gamma Rays Produced Inside the BLR}
\label{insideBLR}

Rapid variability in FSRQs such as 3C 454.3 ($t_v \sim3$\ hours;
\cite{tavecchio10}), and PKS 1510$-$089 ($t_v \sim 1$\ hour;
\cite{brown13,saito13}) limits the size of the emitting region by
Equation (\ref{Rblob}).  If the emitting region takes up the entire
cross section of a conical jet with half-opening angle $\alpha$, then
it should be at a distance
\begin{equation}
r \le 0.1\ \delta_1\ t_{v,4}\ \alpha^{-1}_{-2}\ (1+z)^{-1}\ \pc
\end{equation}
from the base of the jet. Based on scaling relations found from
reverberation mapping, the typical BLR region for FSRQs is
$r_{BLR}\sim 0.1\ \pc\ \approx 10^{17}\ \cm$\ (e.g.,
\cite{bentz06,bentz13}), so that the emitting region would likely be
within the BLR.  An emitting blob inside the BLR would move outside of
a BLR with radius $R_{BLR}$\ in a time period (assuming $\theta \ll
1$)
\begin{equation}
\Delta t = 3.3\times10^4\ R_{BLR,17}\ \delta_1^{-2}\ (\delta/\G)\ (1+z)\ \s
\end{equation}
in the observer's frame.  So if a $\g$-ray flare is found to last
longer than 9 hours or so, it must have moved out of the BLR, or be
made up of many smaller flares from several different blobs.  The
cooling timescale in the observer's frame for electrons producing
photons observed at $E_{obs}=1$\ GeV, assuming a high Compton
dominance and scattering of Lyman $\alpha$ BLR photons
($E_0/(m_ec^2)=2\times10^{-5}$) , is
\begin{equation}
\label{coolBLR}
t_{cool} = 
%\frac{1+z}{\delta}\frac{3 m_e c^2}{4c\sT \G^2 u_{ext}\gp} = 
4400\ u_{ext,-2}^{-1}\ \delta_1^{-2}\ (\delta/\G)^2\ E_{0,Ly\alpha}^{1/2}\ 
E_{{\rm obs},\GeV}^{-1/2}\ (1+z)^{1/2}\ \s
\end{equation}
which implies the electrons can completely cool before the blob moves
out of the BLR.

\subsection{Gamma Rays Produced Outside the BLR}
\label{outsideBLR}

Optical and $\g$-ray flares in FSRQs are often associated with a slow
increase in radio flux which peaks after the $\g$-ray flare
\cite{laeht03,leon11}, and the ejection of superluminal components
from the 43 GHz core (e.g., \cite{jorstad01,marscher08,marscher10}).
According to Marscher \cite{marscher12}, 2/3 of $\g$-ray flares are
associated with the ejection of a superluminal component, from which
the conclusion is drawn that the $\g$-ray flares are coincident with
the 43 GHz core, located at a few pc from the black hole, outside the
BLR (e.g., \cite{marscher10,ramak15}).  There are two arguments that
the 43 GHz core, and thus the $\g$-ray flares occur, at $r>1$\ pc from
the black hole, outside the BLR.
\begin{enumerate}
\item The 43 GHz core has observed radius $R_{core}$ and the jet has
known half-opening angle $\alpha$ from VLBI observations
(e.g.,\cite{jorstad05}).  This gives the distance of the core from the
black hole, $r=R_{core}/\alpha$, which for AO 0235+164, gives a
distance of $r \gtrsim 12$\ pc \cite{agudo11_0235}.
\item The time scales for the radio outbursts are $\Delta t \sim$ 10s
of days.  During this time blobs will have had to travel distances
\begin{equation}
r \approx 1.0\ \Delta t_6\ \delta_1^2\ (\G/\delta)\ (1+z)^{-1}\ \pc
\end{equation}
before the $\g$-ray flares and ejections of the superluminal
components.  Nalewajko et al.\ \cite{nalewajko14} found a possible way
to explain the delay with a light-travel time effect that could allow
$\g$-ray flares to occur at much smaller distances than the location
of the 43 GHz core.
\end{enumerate}
Additionally, $\g$-ray flares with emission up to 100s of GeV have
been detected with imaging atmospheric Cherenkov telescopes from
several FSRQs, including 3C279 \cite{albert08_3c279}, PKS 1510$-$089
\cite{abram13_pks1510,aleksic14_pks1510}, 4C 21.35
\cite{aleksic11_4c21.35,ackermann14_4c21.35}, B0218+357
\cite{mirzoyan14_0218}, and PKS 1441+25
\cite{mirzoyan15_pks1441,mukherjee15_pks1441}.  If these flares
occurred inside the BLR, they would suffer extreme $\g\g$ absorption
(see Section \ref{BLRabsorption}), and these $\g$ rays would not have
been observed.  Flares detected out to several 10s of GeV by the LAT
\cite{pacciani14} are also not likely to originate from inside the BLR
for the same reason.  Several ways have been suggested to avoid $\g\g$
attenuation, such as through transport by neutron beams
\cite{dermer12_neutron} or axions \cite{tavecchio12}.  If rapid
variability ($t_v\sim 10^4\ \s$) does occur at $\gtrsim$ pc scale
distances from the black hole, it would mean the emitting region makes
up a small fraction of the jet cross section, which might be a problem
for a standard ``shock in jet'' model.

Outside the BLR, a likely source of seed photons is the dust torus.  
The region where scattering of dust photons is geometrically likely
is $r<R_{dust}$ where 
\begin{equation}
R_{dust} = 1.1\ L_{disk,45}^{1/2}\ T_3^{-2.6}\ \pc
\end{equation}
is the inner radius of the dust torus (the sublimation radius),
$L_{disk}$ is the disk temperature, and $T$ is the temperature at dust
here \cite{nenkova08_paper1,nenkova08_paper2}.  This constrains
the energy density from the dust torus to be
\begin{equation}
u_{dust} = 2.3\times10^{-5}\ \xi_{-1} T_3^{5.2}\ \erg\ \cm^{-3}
\end{equation}
where $\xi \le 1$ is the fraction of the disk emission reprocessed by 
the dust torus.  An emitting blob with $r<R_{dust}$ moves outside
in a time period
\begin{equation}
\Delta t = 13\ \delta_1^{-2}\ (\delta/\G)\ L_{disk,45}^{1/2}\ T_3^{-2.6}\ (1+z)\ \dday\ .
\end{equation}
The cooling timescale for the electrons making LAT-energy $\g$ rays by
scattering dust torus photons is
\begin{equation}
\label{cooldust}
t_{cool} = 3.5\ \xi_{-1}^{-1}\ T_3^{-4.7} \delta_1^{-2}\ (\delta/\G)^2\ 
E_{{\rm obs},\GeV}^{-1/2} (1+z)^{1/2}\ \dday\ .
\end{equation}
Thus the electrons will have enough time to cool before leaving the
dust torus.  Note that the energy density and cooling timescale are
strongly dependent on the dust temperature, which cannot exceed
$T_3\approx 2$ or else it will sublimate.  The dust torus can only act
as the seed photon source if $r<R_{dust}\sim 1\ \pc$.  However, as
described above, there is evidence that it can exceed this distance,
in which case another seed photon source is needed.  Another
possibility is the scattering of photons from other parts of the jet
\cite{ghisellini05,marscher14,macdonald15}.  Zacharias \& Schlickeiser
\cite{zacharias12} fit several FSRQs with a synchrotron/SSC model
without an EC component, offering another possibility.

\subsection{Gamma Ray Light Curves}
\label{grayLC}

One possible way to determine the seed photon source for EC has been
suggested by Dotson et al.\ \cite{dotson12}.  This would effectively
resolve the question of the location of the $\g$-ray emitting region
(Section \ref{location}), since if the region is withing the BLR, the
seed photons will be at a much higher energy ($\approx 10$\ eV for
Ly$\alpha$) than if they are from a $\approx 1000$\ K blackbody.  This
involves exploiting the Klein-Nishina cross section.  Due to the
turnover in the Compton-scattering cross section at high energies, the
relative cooling of electrons producing $\g$ rays at different
energies will be different for different seed photon energies, which
could be used to constrain the seed photon energy.  This requires
using at least two LAT light curves with two different energy ranges.
The approximate timescales of the flares needed can be found from
Equations (\ref{coolBLR}) and (\ref{cooldust}).  Application
of this technique to several bright flares from PKS 1510$-$089 
observed in 2009 indicates that some flares are produced 
inside the BLR, while some are produced outside \cite{dotson15}.

\subsection{Gravitational Lensing}

Gravitational lensing of a blazar by an intervening galaxy can produce
multiple images of the blazar.  The images will have different
brightnesses and the light curve of one will be delayed with respect
to the other.  Although the images can often be resolved with radio
telescopes, this is not possible with the larger PSF of $\g$-ray
telescopes.  However, the differing brightness and time delay allow
for the possibility that one can distinguish the $\g$-ray light curves
of two lensed images, particularly during bright flares.  Barnacka et
al.\ \cite{barnacka11} claimed the first detection of a
gravitationally lensed time delay in $\g$-rays in the LAT light curve
of the FSRQ PKS 1830$-$211 (although see \cite{abdo15_1830}) and
another $\g$-ray gravitational lens-induced delay was found from the
blazar B0218+357 \cite{cheung14}.  For some of the bright $\g$-ray
flares reported in the literature thus far, either the magnification
ratio (the ratio of the brightness of two lensed images), or the time
delay, or both, are not consistent with the ones measured from the
radio images.  This has been interpreted in two ways.

One interpretation is that the location of the emitting region is
different for $\g$-ray flares than most of the radio emission
\cite{barnacka14}.  In this case, carefully modeling the lens system
can constrain the location of the $\g$-ray emitting region in the jet.
For two $\g$-ray flares detected from PKS 1830$-$211, the location of
the flares were constrained to originate at $\gtrsim 1$\ kpc projected
distance from the radio core, or $\gtrsim 10$\ kpc deprojecting with a
jet angle to the line of sight of $\theta=0.1$\ rad \cite{barnacka15}.
This is far beyond the BLR and dust torus, so that it is not clear
what the seed photon source would be for EC at these large distances.

Another possible interpretation is microlensing by individual stars in
the lensing galaxy.  If this is the case, the size of the $\g$-ray
emitting region can be constrained.  This has been used to constrain
it to be $R\sim 10^{14}$\ cm for flares in both PKS 1830$-$211
\cite{neronov15} and B0218+357 \cite{vovk15}.  Assuming the emitting
region takes up this entire cross section of the jet (see Section
\ref{insideBLR}), this constrains its location to be $r\sim10^{16}\
\alpha_{-2}^{-1}\ \cm$\ from the base of the jet.  The blazar
B0218+357 has been detected at $>100$\ GeV by MAGIC
\cite{mirzoyan14_0218}.  It is not clear if this source has a
significant BLR, and it is classified as a ``blazar of uncertain
type'' in BZCAT\footnote{http://www.asdc.asi.it/bzcat/}, but if it
does, it is not clear how $\g$-ray photons would escape avoiding
$\g\g$ absorption with BLR photons (see Section \ref{outsideBLR}).

\section{Variability}
\label{variability}

One of the defining characteristics of blazars is extreme variability
across the electromagnetic spectrum, on timescales as short as minutes
in some cases.  Here I briefly describe a few results from modeling
blazar variability.

\subsection{SEDs at Different Epochs}

A number of FSRQs have been modeled with time-independent
synchrotron/SSC/EC models in several states, and an interesting pattern
is emerging.  For many sources, the quiescent and flaring states can
be modeled by changing only the electron distribution between
states--that is, the other parameters, magnetic field, Doppler factor,
etc., are not changed.  This includes flares from PKS 0537$-$441
\cite{dammando13_0537} (see Figure \ref{0537sed}), PKS 1830$-$211
\cite{abdo15_1830}, and 4C 21.35 \cite{ackermann14_4c21.35}.  For
other sources, different flares can be modeled by changing at least
one additional parameter (usually the magnetic field), while for other
flares from the same sources, only changes in the electron
distribution are needed.  These include PKS 2142$-$75 \cite{dutka13}
and PKS 1424$-$418 \cite{buson14}.  Flares where another parameter is
needed are usually those where $\g$-ray flares occur without
corresponding optical flares, or optical flares without corresponding
$\g$-ray flares \cite{chatterjee13,cohen14}.

\subsection{Multi-Wavelength Light Curves}

Recently, several authors have begun modeling the multi-wavelength
light curves of blazars with time-dependent leptonic models.  Saito et
al.\ \cite{saito15} have modeled a bright LAT-detected $\g$-ray flare
from PKS 1510$-$089 with duration $\sim$ a few hours in 2011 November.
Unfortunately, light curves at other wavelengths are not available for
this flare.  Using a model of a shock traveling through a helical
magnetic field, Zhang et al.\ \cite{zhang15} were able to reproduce
the light curves, including polarization angle and degree as a
function of time, for a flare from 3C 279 with duration $\sim20$\
days.

\subsection{Fourier Analysis}

Theoretical work has also been done on reproducing the power spectral
densities of blazars.  A one-zone model was created where variability
is caused by only variability in the electron injection rate, assumed
to be injected stochastically as as a power-law in Fourier frequency
(i.e., red noise) with index $a$.  The model predicts that at low
frequencies, synchrotron and EC PSDs would be $S(f)\propto f^{-a}$,
while SSC PSDs would be $S(f)\propto f^{-(2a-2)}$ \cite{finke14}.
This predicts a separation in the index in the LAT PSDs of BL Lacs and
FSRQs, assuming BL Lacs emit $\g$ rays by SSC and FSRQs emit $\g$ rays
by EC.  The predicted separation was in fact seen in the LAT PSDs
calculated by Nakagawa \& Mori \cite{nakagawa13}, however, not in the
LAT PSDs calculated by Sobolewska et al.\ \cite{sobolewska14} or
Ramakrishnan et al.\ \cite{ramak15}.

At higher frequencies, for synchrotron and EC, one expects breaks in
the PSDs of blazars, so that the power-law index will change from $a$
to $a+2$.  The frequency of the breaks will be related to the cooling
timescale.  Finding this break in the $\g$-ray PSDs at several LAT
energies could be used to constrain the seed photon energy
\cite{finke15}, similar to the method described by Dotson et al.\
\cite{dotson12} (see Section \ref{grayLC}).

\section{Hadronic Models}
\label{hadronic}

In the {\em Fermi} era, several authors have fit the SEDs of
LAT-detected blazars with hadronic models.  Here we briefly review a
few of the results from hadronic modeling.

B\"ottcher et al.\ \cite{boett13} have modeled the SEDs of a dozen
blazars, all of which are LSP or ISP type and half of which are FSRQs,
with the synchrotron proton blazar model \cite{muecke01,muecke03}.  In
many cases the models implied jet powers of $P_j \sim 100-1000
L_{Edd}$, which seems to strongly disfavor these models
\cite{zdziarski15}.  This is consistent with the result of Petropoulou
\& Dimitrakoudis \cite{petro15_3c273}, who find that the synchrotron
proton model is disfavored for the FSRQ 3C 273 due to unreasonably
high magnetic field and jet power values.  However, several HSP BL
Lacs were fit with a similar hadronic model by Cerruti et al.\
\cite{cerruti15}, which gave more reasonable jet powers.  Petropoulou
\& Mastichiadis \cite{petro15_pair} show that a signature of hadronic
models of BL Lac objects is a bump at $\approx 40$\ keV to 40 MeV from
synchrotron emission of pairs produced by the Bethe-Heitler process.
Such a feature may be detectable by future $\g$-ray telescopes
sensitive in this range.  Aside from energetics, leptonic and hadronic
models could be distinguished through the detection of a very-high
energy neutrino by IceCube or another such detector coincident in
direction and time with a $\g$-ray flare
\cite{diltz15,petro15_neutrino}, or through polarization of X-rays or
$\g$-rays \cite{zhang13_polar}.  Several authors have recently
developed time-dependent hadronic models
\cite{dimitrakoudis12,diltz15}, which could offer an additional and
probably more observationally realistic way to discriminate between
leptonic and hadronic models.

If blazars are the source of UHECRs, then it is possible that cosmic
rays that escape the blazar itself could make an observational
signature through pion production and decay through p$\g$ interactions
with the EBL and CMB.  This could explain the non-variable VHE
emission for HSP BL Lacs \cite{essey10_1,essey10_2,essey11_cr}.

\section{Acknowledgments}

I am grateful to Soebur Razzaque for the invitation to visit the
University of Johannesburg and to attend and speak at the HEASA
meeting, and to Andrew Chen for the opportunity to visit the
University of Witwatersrand.  I am also grateful to Markus B\"ottcher
for the additional opportunity to visit his group in Potchefstroom.
Charles Dermer, Teddy Cheung, and Anna Barnacka have my thanks for
proofreading portions of this manuscript (any mistakes are soley my
responsibility).  Some of my research presented here was funded by the
Chief of Naval Research and by NASA {\em Fermi} Guest Investigator
grants.

%\bibliographystyle{siam}
%\bibliography{3c454.3_ref,EBL_ref,references,mypapers_ref,blazar_ref,sequence_ref,SSC_ref,LAT_ref,variability_ref,gravlens_ref}

\end{document}